\documentclass[sigconf]{aamas}

\usepackage{amsmath}
\usepackage{booktabs}
\usepackage{multirow}
\usepackage{xspace}
\usepackage{balance}
\usepackage{tabularx}
\usepackage{array}

\newcolumntype{Y}{>{\raggedright\arraybackslash}X}

\title{Amadeus: When Models of People Meet}

\author{Karl Hanna}
\affiliation{
  \institution{Queen's University Belfast}
  \country{United Kingdom}
}

\acmConference[AAMAS 2027]{Submitted to AAMAS 2027}
{3--7 May 2027}{Hanoi, Vietnam}

\setcopyright{none}

\acmISBN{}
\acmDOI{}


\newcommand{\GG}{\textsc{GG}\xspace}

\newcommand{\AB}{\textsc{AB}\xspace}

\begin{document}

\begin{abstract}

With the constant advancements in AI, one possibility is to model agents after humans and, in turn, use these agents to carry out synthetic interactions. Such models could be used to predict interactions between their real counterparts, or potentially interactions at larger scales. In this paper, we test a more controlled version of this question through chess. We use 8 elite chess players, seal their direct pairwise games, learn each player independently using different methods, and then compose the resulting models on the withheld dyads. To evaluate the generated interactions, we use two measurements: opening-family total variation distance and win-draw-loss (WDL) total variation distance. M1 primarily improves WDL fidelity while producing smaller opening-family improvements, whereas M2 produces much larger opening-family improvements while having little effect on WDL-TV. For opening-family behaviour under M2, the correct assignment of the eight learned player identities also gives the closest match among all \(8! = 40{,}320\) possible assignments. These results show that at least some properties of previously unseen interactions can be recovered from independently learned individuals. The partial recovery observed here may reflect limitations of the current individual modelling methods rather than a fundamental limit on compositional interaction recovery. An additional post-hoc method that combines the two mechanisms improves both measurements, suggesting that recovery across these behavioural properties is not necessarily mutually exclusive.

\end{abstract}

\maketitle

\section{Introduction}
\label{sec:introduction}

With the rapid progress in machine learning and AI in recent years, it is now increasingly possible to model agents after humans. This has been explored in different applications, including modelling individual behaviour \cite{omi2025generative,park2026selfreports} and identifying or reproducing a chess player's style \cite{mcilroyyoung2021stylometry,mcilroyyoung2022individual,sogliuzzo2026playerspecific}.

One specific question remains less explored. With the premise that we have two agents modelled accurately after real humans, what happens if we have them interact together? Individual behaviour may depend on things such as the person being interacted with, relationships between individuals, and other interaction-specific effects, meaning that individual fidelity does not necessarily imply interaction fidelity. However, if the composition between individual models holds, then this could enable simulations involving interacting models of real individuals, small teams, and even larger social systems.

In this paper, we test this question. For a cohort of target players, we learn each player's behaviour using only games against players outside the target cohort. All interactions between target players are withheld. We then compose pairs of learned player models and compare their synthetic interactions with the corresponding sealed real interactions.

We use chess as the domain for several reasons. The interaction between two players is pairwise and explicit, their actions are discrete and completely observable, and the environment and rules are fixed. Chess also provides repeated observations of each player and repeated interactions between the same pairs, while allowing all interactions between a target pair to be cleanly withheld to prevent leakage. Outcomes and behavioural properties, such as the openings played, can also be quantitatively measured. Finally, unlike something more open-ended such as conversation, there is little ambiguity about what the interaction trajectory actually was.

We use Amadeus to test this. Amadeus uses a ChessFormer-based model trained from scratch, which allows us to control the data seen by the base model, rather than using an already-trained model and risking leakage from games between the eight elite target players. We then use two different methods to model specific players while keeping all interactions between the target players sealed from training: Policy-Level Personalization (M1) and Candidate-Level Personalization (M2). These player models are composed in synthetic games and compared with the real interactions that were withheld. After evaluating M1 and M2, we also explore a hybrid (M3) that combines their already-trained mechanisms without additional training.

We find that compositional recovery is aspect, and method-dependent: M1 more clearly improves outcome fidelity, while M2 produces much larger improvements in opening-family fidelity. For M2 and the exploratory M3 hybrid, the correct assignment of player identities gives the best opening-family match among all \(8! = 40{,}320\) possible assignments, showing that the effect depends on the specific learned identities rather than personalization alone.

Our main contributions are as follows:
\begin{itemize}
    \item We study whether independently learned models of specific individuals can reproduce their real interactions when those interactions are completely withheld during training.
    \item We introduce Amadeus, a controlled chess-based setup using eight elite players and strict pairwise sealing to test this question.
    \item We evaluate two different methods for modelling player-specific behaviour and compare their synthetic interactions against withheld real interactions.
\end{itemize}
\section{Related Work}
\label{sec:related-work}

Chess has increasingly been used to model human rather than optimal decision-making. Maia trains policies on human games to predict moves at different skill levels \cite{mcilroyyoung2020maia}, while Maia-2 unifies behavior across Elo levels within a single skill-conditioned model \cite{tang2024maia2}. Subsequent work has moved from population-level behavior toward specific individuals. Per-player fine-tuning improves prediction of individual players \cite{mcilroyyoung2022individual}, while behavioral stylometry shows that player identity can be recovered from decision traces alone \cite{mcilroyyoung2021stylometry}. More recent approaches improve the efficiency or representation of personalization: Maia4All uses prototype-based initialization to learn individual behavior from substantially fewer games \cite{tang2026maia4all}, while champion-specific embeddings have been combined with distributional style evaluation based on Jensen--Shannon divergence \cite{sogliuzzo2026playerspecific}. Matilda instead combines a frozen human policy with independent search evidence and lightweight player-style embeddings \cite{carlson2026matilda}. Our base architecture, ChessFormer, provides the Maia-3 human-move models and reports 57.1\% move-matching accuracy \cite{monroe2026chessformer}.

Individual-level behavioral modeling also extends beyond chess. Omi et al.\ formulate individual behavior generation as a multi-task learning problem, learning compact person-specific style vectors within a shared model at scale across chess and Rocket League \cite{omi2025generative}. In a substantially different setting, Park et al.\ construct LLM agents representing 1,052 real individuals from interview and survey self-reports and evaluate them on held-out attitudes, personality measures, economic-game behavior, and experimental responses \cite{park2026selfreports}. Their combined interview-and-survey agents reach performance equivalent to 86\% of participants' own two-week test--retest consistency on held-out survey questions. Together, these results illustrate that models can capture aspects of specific individuals across different domains and forms of behavioral evidence.

A parallel line of work studies interactions among generative agents. Generative Agents places multiple language-model agents in a shared environment and demonstrates emergent behaviors including information diffusion, relationship formation, and coordination \cite{park2023generativeagents}. SOTOPIA provides an environment for evaluating agents across social interactions such as cooperation, negotiation, and competition \cite{zhou2024sotopia}, while Concordia provides infrastructure for constructing grounded generative agent-based simulations \cite{vezhnevets2023concordia}. These works study believable, socially capable, or emergent interaction among agents, rather than whether interactions between models of specific real individuals reproduce those individuals' real interactions. In contrast, Amadeus withholds all interactions among a target cohort during learning, composes the resulting individual models afterward, and evaluates the synthetic interactions against the corresponding withheld real interactions.
\section{Methodology}
\label{sec:methodology}

\subsection{Problem Formulation}

For target players \(A\) and \(B\), let \(\pi_A\) and \(\pi_B\) denote the learned behavioral policies for the two players, trained without access to direct \(A\)--\(B\) interactions. Composing these policies in the shared chess environment induces a synthetic interaction distribution,

\begin{equation}
    \widehat{P}_{AB}
    =
    \mathcal{C}(\pi_A,\pi_B),
\end{equation}

where \(\mathcal{C}\) denotes the interaction process between the two policies. Let \(P^{\mathrm{real}}_{AB}\) denote the empirical distribution of the sealed real interactions between \(A\) and \(B\). Our goal is to test whether the synthetic interaction distribution \(\widehat{P}_{AB}\) resembles \(P^{\mathrm{real}}_{AB}\).

\subsection{Base Chess Model}

Both personalization methods use a base model based on ChessFormer
\cite{monroe2026chessformer}.

The board is first canonicalized and encoded as a tensor of shape \(m \times 64 \times 96\), where \(m\) is the batch size. Each square contains eight groups of 12 one-hot features, representing the current board and the previous seven board states.

The model also receives two 128-dimensional Elo embeddings, one for the mover and one for the opponent. Each embedding is obtained by interpolating between learned low- and high-Elo vectors. These embeddings are broadcast across the board and concatenated with each square's piece and history features, giving a 352-dimensional input per square.

Each square's input is projected to \(d_{\mathrm{model}} = 1024\) and passed through eight ChessFormer blocks. Each block contains geometric attention bias (GAB) augmented multi-head self-attention followed by a feed-forward network.

To describe the GAB computation, let \(N\) denote the number of board squares, \(h\) the number of attention heads, and \(d_1\), \(d_2\), and \(d_3\) the GAB projection dimensions. The block first projects each square representation from \(d_{\mathrm{model}}\) to \(d_1\). The resulting \(N \times d_1\) representation is flattened into a vector of dimension \(Nd_1\), then projected to \(d_2\), followed by GELU and LayerNorm. A further projection maps this vector to \(hd_3\), again followed by GELU and LayerNorm. The result is reshaped into \(h \times d_3\), giving one latent vector per attention head.

The same linear map is applied to each head's vector, projecting it from \(d_3\) to \(N^2\). Reshaping these outputs produces one \(N \times N\) GAB bias matrix per head. Each bias matrix is added to the corresponding standard attention logits before softmax. The resulting attention weights are then multiplied by the value representations, as in standard multi-head self-attention.

We use \(N = 64\), \(d_{\mathrm{model}} = 1024\), \(h = 32\), \(d_1 = 32\), and \(d_2 = d_3 = 128\).

After the final block, the square representations are passed to policy and value heads. The policy head uses separate projections to represent each square as a possible source (\texttt{FROM}) and destination (\texttt{TO}). Scaled pairwise dot products between these representations produce \(64 \times 64 = 4096\) source--destination logits. Promotions are handled separately because the same source and destination can correspond to different promotion pieces, adding another 256 logits. The full policy output therefore contains 4352 logits.

The value head mean-pools the square representations and passes the resulting vector through an MLP with dimensions \(1024 \rightarrow 1024 \rightarrow 3\), producing three outcome logits.

We train the base model from scratch because the publicly released weights may have been trained on games involving our target players.

Let \(x_t\) denote the encoded board history immediately before ply \(t\),
and let \(e(r)\in\mathbb{R}^{128}\) denote the base model's Elo embedding
for rating \(r\). We write the base policy logits as

\begin{equation}
    \ell_t
    =
    f_\theta\!\left(
        x_t,
        e(r_t),
        e(r_t^{\mathrm{opp}})
    \right)
    \in \mathbb{R}^{4352},
\end{equation}

where \(r_t\) and \(r_t^{\mathrm{opp}}\) are the mover's and opponent's
ratings at ply \(t\), respectively.

\subsection{Target Cohort and Interaction Sealing}

Our target cohort contains eight players: Magnus Carlsen, Hikaru Nakamura, Ian Nepomniachtchi, Wesley So, Fabiano Caruana, Maxime Vachier-Lagrave, Levon Aronian, and Alireza Firouzja. These players were selected to provide dense mutual-game coverage across all 28 target-player pairs.

Base-model training consists of two stages. In the first stage, we train on public Lichess games from June of each year from 2017 to 2021. Using the same calendar month across years is intended to limit seasonal variation in the training data.

We exclude games involving any identified Lichess account associated with a target player, including accounts with low-confidence attribution, to reduce the risk of leakage. 

Games are processed in chunks of 20,000 and assigned to Elo bins using the mean of White's and Black's ratings. We use 22 bins: one below 600, twenty 100-point bins covering \([600,2600)\), and one for ratings of 2600 and above. Within each chunk, we retain at most 10 games per bin. The counts reset for the next chunk, and processing of a chunk can stop early once all bins are full. This sampling is intended to balance representation across playing strengths rather than concentrate training on the middle of the Elo distribution.

In the second stage, we fine-tune the base model for an additional 200{,}000 steps on the Lichess Broadcast corpus \cite{lichessbroadcast}. We exclude every Broadcast game in which either player is one of the eight targets.

For both personalization methods, each target player's eligible data consists only of games against players outside the target cohort. These games are split at the game level for training and individual behavioral validation. All games between target players remain sealed and are excluded from training, tuning, and calibration.

\subsection{Individual Behavioral Modeling}

We use two personalization methods. Policy-Level Personalization changes the player conditioning supplied to the base model, while Candidate-Level Personalization adds player-specific adjustments to candidate move logits.

\subsubsection{Policy-Level Personalization (M1)}

M1 adapts the player personalization approach of Sogliuzzo et al. ~\cite{sogliuzzo2026playerspecific}, which learns a player-specific embedding while keeping the Maia-2 backbone frozen, to ChessFormer's continuous Elo conditioning.

For each target player \(p\), M1 introduces a trainable player vector \(z_p \in \mathbb{R}^{128}\), initialized from the base model's Elo embedding at that player's representative Elo,

\begin{equation}
    z_p^{(0)} = e(r_p).
\end{equation}

The base-model parameters \(\theta\) remain frozen. When player \(p\) is the mover at ply \(t\), the ordinary mover-Elo embedding is replaced by \(z_p\), giving personalized policy logits

\begin{equation}
    \ell_t^{(p)}
    =
    f_\theta\!\left(
        x_t,
        z_p,
        e(r_t^{\mathrm{opp}})
    \right).
\end{equation}

The opponent's Elo conditioning remains unchanged, and \(z_p\) is the only parameter optimized for that player.

During generation, illegal moves are masked before softmax, giving a probability distribution over legal moves. Because the base model was initially trained on a broad range of playing strengths and is used here to model elite grandmasters, we apply a Stockfish-based sampling adjustment.

\paragraph{Production sampling.}
For generation, illegal moves are first masked and the calling behavioral policy is restricted to its top-\(K\) legal candidates, \(\mathcal{C}_t=\{a_{t,1},\ldots,a_{t,K}\}\), with \(K=5\). Let \(p_{t,i}\) denote the behavioral-policy probability of candidate \(a_{t,i}\), and let \(c_{t,i}\ge 0\) denote its centipawn loss relative to the best candidate in \(\mathcal{C}_t\), estimated by Stockfish~19 at depth~8.
\begin{equation}
    q_{t,i}
    =
    \operatorname{softmax}_i
    \left(
        \log p_{t,i}
        -
        \lambda\frac{c_{t,i}}{100}
    \right),
    \qquad
    \lambda = 2.
\end{equation}

The coefficient \(\lambda\) is frozen using nonsealed data and is chosen to limit intervention in the behavioral distribution while discouraging poor moves. The same sampling rule is used for the generic model, M1, M2, and the exploratory M3 hybrid.

\subsubsection{Candidate-Level Personalization (M2)}

M2 begins from the top-\(K\) legal candidates of the frozen base policy, with \(K=5\). Let \(\mathcal{C}_t=\{a_{t,1},\ldots,a_{t,K}\}\) denote this candidate set at ply \(t\), and let \(\ell_{t,i}\) denote the frozen base-policy logit of candidate \(a_{t,i}\).

A shared move-style network \(g_\psi\) encodes the current board and candidate move as a \(d\)-dimensional feature vector,

\begin{equation}
    \phi_{t,i}
    =
    g_\psi(x_t,a_{t,i})
    \in \mathbb{R}^{d},
    \qquad d=32.
\end{equation}

Each target player \(p\) has a learned style vector \(u_p\in\mathbb{R}^{d}\). The candidate and player representations are L2-normalized,

\begin{equation}
    \widehat{\phi}_{t,i}
    =
    \frac{\phi_{t,i}}{\|\phi_{t,i}\|_2},
    \qquad
    \widehat{u}_p
    =
    \frac{u_p}{\|u_p\|_2}.
\end{equation}

M2 adds a player-specific residual to the frozen base-policy logit,

\begin{equation}
    \tilde{\ell}_{t,i}^{(p)}
    =
    \ell_{t,i}
    +
    \sigma\sqrt{d}\,
    \widehat{\phi}_{t,i}^{\top}\widehat{u}_p,
\end{equation}

where \(\sigma\) is a learned scalar shared across players and candidates. The shared move-style parameters \(\psi\), player vectors \(u_p\), and \(\sigma\) are trained jointly across all eight target players, while the base-model parameters \(\theta\) remain frozen.

During training, if the observed human move is absent from the raw top-\(K\) candidate set, it is appended so that the objective is defined. At inference, the candidate set remains the raw top-\(K\). Softmax over the adjusted logits gives the M2 behavioral probabilities, which are passed through the same production sampling rule defined for M1.

\subsection{Individual Behavioral Validation}

Before evaluating composed interactions, we test whether each of the two personalization methods learns player-specific behavioral information. This validation uses only held-out target-vs-outsider games, with game-level train-validation splits. No sealed target-target interactions are used.

For each target player, we evaluate their held-out moves using the correct learned player representation, the generic model, and the representations of the other seven target players. We compare the negative log likelihood assigned to the observed human move. A lower negative log likelihood means that the model assigns greater probability to the move the player actually made.

We also rank all eight learned player representations by their likelihood on each player's held-out data. This tests whether the correct representation is more consistent with that player's behavior than the representations learned for other players.

These likelihoods are measured directly from the behavioral models, before and independently of the Stockfish sampling adjustment. For M1, validation is performed over the full legal action space. For M2, the generic, correct-player, and wrong-player comparisons use the same candidate set: the raw top-five legal moves from the frozen base policy, with the observed human move appended if it is absent.

We separately measure raw top-five coverage before appending the observed move. This quantifies how often the human move is already available to the candidate-level personalization mechanism.

\subsection{Exploratory Hybrid Composition (M3)}

After inspecting the M1/M2 interaction results, we introduce M3 as a post-hoc exploratory hybrid of the two personalization mechanisms. M3 uses the already-trained M1 and M2 components without additional training or optimization.

The M1-personalized policy produces the top-\(K\) legal candidate moves. The already-trained M2 candidate-level mechanism then reranks these candidates by adding its player-specific residual to their M1-personalized policy logits. The resulting candidates are passed to the same production sampler used by M1 and M2. All learned components remain frozen, and no parameters or sampling settings are tuned on sealed interactions.

\subsection{Compositional Interaction Protocol}

The primary sealed-dyad evaluation compares M1 and M2. We also apply the same protocol to M3 as a post-hoc exploratory evaluation.

The eight target players form 28 unordered dyads. For a target player \(p\), let \(\pi_p^G\) denote the generic generation policy and let \(\pi_p^M\) denote the personalized generation policy under method \(M\), where

\[
    M \in \{\mathrm{M1}, \mathrm{M2}, \mathrm{M3}\}.
\]

For each dyad \((A,B)\), the four experimental conditions produce the following synthetic interaction distributions:

\begin{align}
    \widehat{P}_{AB}^{GG,M}
    &=
    \mathcal{C}\!\left(\pi_A^G,\pi_B^G\right),
    \\
    \widehat{P}_{AB}^{AG,M}
    &=
    \mathcal{C}\!\left(\pi_A^M,\pi_B^G\right),
    \\
    \widehat{P}_{AB}^{GB,M}
    &=
    \mathcal{C}\!\left(\pi_A^G,\pi_B^M\right),
    \\
    \widehat{P}_{AB}^{AB,M}
    &=
    \mathcal{C}\!\left(\pi_A^M,\pi_B^M\right).
\end{align}

Thus, \textbf{GG} uses generic models for both players, \textbf{AG} personalizes only \(A\), \textbf{GB} personalizes only \(B\), and \textbf{AB} personalizes both players. The generic baseline is not a low-skill or population-only model: it is the same frozen base after adaptation to elite Broadcast games, with all the target players excluded. The labels identify the modeled players rather than their colors. Each condition is compared with the same sealed empirical distribution \(P_{AB}^{\mathrm{real}}\).

For each dyad and condition, we generate 10,000 games: 5,000 with the \(A\) side playing White and the \(B\) side playing Black, and 5,000 with the colors reversed. Across 28 dyads and four conditions, this gives 1.12 million games per method.

Games use normal rule-based termination, including claimable draws by threefold repetition and the fifty-move rule, implemented through \texttt{board.outcome(claim\_draw=True)}. We do not model voluntary human actions such as resignation or agreed draws. An additional emergency limit is set at 500 plies. Games reaching this limit without a rule-based outcome are censored rather than counted as draws.

\subsection{Evaluation Metrics}

We use two main metrics to compare synthetic interactions with the sealed real games: total variation distance between win--draw--loss (WDL) distributions and total variation distance between opening-family distributions. For both metrics, we use total variation distance (TV) to be able to compare across different dyads and generation methods. The evaluation uses 1609 sealed games covering all 28 dyads, with 25--131 real games per dyad.

\begin{equation}
    \mathrm{TV}(P,Q)
    =
    \frac{1}{2}
    \sum_x
    \left|P(x)-Q(x)\right|.
\end{equation}

For WDL evaluation, outcomes are defined by player identity: an \(A\) win, a draw, or a \(B\) win. The two color orientations are evaluated separately, and their WDL total variation distances are combined with equal 50/50 weight. Censored 500-ply games do not enter the WDL denominator and are reported separately rather than being treated as draws.

For opening-family evaluation, games are classified using a pinned revision of the \texttt{lichess-org/chess-openings} database \cite{lichesschessopenings}. Each game is replayed, and its deepest recognized opening position is used to assign an opening name. The full name is then collapsed to its top-level family, generally using the part before the first colon. For example, a named Sicilian variation is assigned to the \textit{Sicilian Defense} family. Games with no recognized opening are assigned an explicit \texttt{Unknown} family rather than discarded.

We compare opening-family frequencies in the generated games with those in the sealed real games using total variation distance. For both metrics, lower distance indicates closer agreement with the sealed real distribution for the property being measured.

\subsection{Post-hoc robustness analyses}

After inspecting the primary interaction results, we conducted two additional analyses to probe their robustness and interpretation. First, because the sealed interaction sets contain only 25--131 real games per dyad, we estimate a model-conditioned finite-sample simulation to quantify how much TV can arise from the limited historical sample alone. For each dyad and color assignment, we treat the fixed synthetic distribution as the reference distribution and draw 10,000 samples matching the corresponding number of real games, recomputing TV for each replicate. 

Second, to test whether the observed recovery depends on composing the correct player identities rather than reflecting a cohort-wide personalization effect, we compare the matched compositions against mismatched identities. We evaluate all \(8! = 40{,}320\) global identity permutations and additionally perform one-player substitution controls in which one member of each dyad is held fixed while the other is replaced by each remaining target player.
\section{Results}
\label{sec:results}

\subsection{Individual Behavioural Validation}
\label{sec:results_individual}

Before evaluating sealed interactions, we first perform a sanity check that the learned representations capture player-specific behaviour. For each of the eight target players, we evaluate held-out nonsealed positions using the correct player representation, the generic representation, and the representations of the seven other target players.

For M1, we evaluate the legal-masked full-action policy negative log-likelihood (NLL). For M2, whose personalized scorer operates over a candidate set, we instead use candidate-matched NLL over the raw top-5 candidates, appending the observed move when it is absent. The NLL values of M1 and M2 therefore have different definitions and should not be directly compared across methods.

For both M1 and M2, the correct player representation achieves the lowest NLL of all eight player representations for every target player. The correct representation ranks first for 8/8 players under both methods. Relative to the generic representation, the mean NLL decreases by 0.006802 for M1 and 0.009278 for M2. Relative to the mean of the incorrect player representations, the corresponding gaps are 0.004365 and 0.020449, respectively. M2's raw top-5 candidate set contains the observed move for 94.98\% of positions on average.

\begin{table}[t]
\centering
\caption{Individual behavioural validation on held-out nonsealed data. $\Delta$NLL denotes generic-minus-correct NLL. M2 uses candidate-matched NLL and is therefore not directly comparable in scale with M1.}
\label{tab:individual-validation}
\small
\begin{tabular}{lrrrr}
\toprule
Method & Rank 1 & Mean $\Delta$NLL & Wrong$-$correct & Positions \\
\midrule
M1 & 8/8 & 0.006802 & 0.004365 & 47,136 \\
M2 & 8/8 & 0.009278 & 0.020449 & 46,950 \\
\bottomrule
\end{tabular}
\end{table}

\subsection{Sealed Pairwise Interaction Fidelity}
\label{sec:results_pairwise}

We next evaluate the central question of the paper: whether the independently learned player models reproduce real interactions that were completely sealed from personalization. We compare the four conditions GG, AG, GB, and AB over all 28 target-player dyads. Results are averaged equally across dyads and across both color orientations. For both WDL-TV and opening-family TV, lower values indicate that the synthetic interaction distribution is closer to the withheld real interaction distribution.

\begin{table*}[t]
\centering
\caption{Sealed target--target interaction fidelity for M1 and M2. Lower TV is better. The final column reports $AB-GG$, so negative values indicate that composing both personalized players reduces the distance from the corresponding real interaction.}
\label{tab:pairwise-main}
\small
\begin{tabular}{llrrrrr}
\toprule
Method & Metric & GG & AG & GB & AB & $AB-GG$ \\
\midrule
M1 & WDL-TV & 0.164171 & 0.157753 & 0.156141 & 0.150148 & -0.014024 \\
M1 & Opening-family TV & 0.554433 & 0.547406 & 0.544765 & 0.536973 & -0.017461 \\
\midrule
M2 & WDL-TV & 0.163166 & 0.162582 & 0.162589 & 0.163408 & 0.000241 \\
M2 & Opening-family TV & 0.554711 & 0.501370 & 0.488343 & 0.437202 & -0.117509 \\
\bottomrule
\end{tabular}
\end{table*}

The two personalization mechanisms produce notably different patterns. For M1, personalization progressively reduces WDL-TV from 0.164171 under GG to 0.150148 under AB, a reduction of 0.014024. Personalizing only one side also reduces the distance, with AG and GB reaching 0.157753 and 0.156141, respectively. M1 also improves opening-family fidelity, although by a smaller amount: opening-family TV decreases from 0.554433 under GG to 0.536973 under AB.

M2 shows almost the opposite pattern across the two behavioural measures. WDL-TV remains effectively unchanged, moving from 0.163166 under GG to 0.163408 under AB. In contrast, opening-family TV decreases substantially. Personalizing a single player reduces it to 0.501370 under AG and 0.488343 under GB, while composing both personalized players reduces it further to 0.437202. This corresponds to an $AB-GG$ change of -0.117509.

These results suggest that the two personalization mechanisms transfer different aspects of player-specific behaviour into unseen interactions. M1 more clearly changes the distribution of game outcomes, whereas M2 produces a much larger change in opening-family behaviour.

\subsection{Exploratory Hybrid Model (M3)}
\label{sec:results_m3}

The complementary behaviour of M1 and M2 motivates an exploratory hybrid, M3, which combines the already-trained M1 player vectors with the already-trained M2 candidate reranker. M3 introduces no additional training or new scientific hyperparameters, and we therefore treat it as a post-hoc exploratory analysis rather than a third primary method.

\begin{table}[t]
\centering
\caption{Exploratory M3 results. The M1 and M2 columns show the corresponding fully personalized AB conditions for reference. Lower TV is better.}
\label{tab:m3}
\small
\begin{tabular}{lrrr}
\toprule
Metric & M1 AB & M2 AB & M3 AB \\
\midrule
WDL-TV & 0.150148 & 0.163408 & 0.152036 \\
Opening-family TV & 0.536973 & 0.437202 & 0.444428 \\
\bottomrule
\end{tabular}
\end{table}

M3 reduces WDL-TV from its own GG value of 0.163073 to 0.152036 under AB, for a change of -0.011037. Its AB WDL-TV is therefore close to the M1 value of 0.150148 and substantially lower than the M2 value of 0.163408.

For opening behaviour, M3 reduces opening-family TV from 0.554395 under GG to 0.444428 under AB, for a change of -0.109967. This is close to the M2 AB value of 0.437202 and substantially below the M1 value of 0.536973. Thus, descriptively, the hybrid retains much of the M1-like improvement in outcome fidelity while also retaining much of the M2-like improvement in opening-family fidelity.

\subsection{Dyad-Level Variation}
\label{sec:results_dyads}

The aggregate results can obscure whether the observed changes are consistent across player pairs. We therefore examine $AB-GG$ separately for each of the 28 dyads. Negative values indicate that personalization brings the synthetic interaction closer to the real interaction.

\begin{table*}[t]
\centering
\caption{Variation in $AB-GG$ across the 28 target-player dyads. ``Improved'' reports the number of dyads for which AB has lower TV than GG.}
\label{tab:dyad-summary}
\small
\begin{tabular}{llrrrrr}
\toprule
Method & Metric & Improved & Mean & Median & Min & Max \\
\midrule
M1 & WDL-TV & 22/28 & -0.014024 & -0.010900 & -0.088700 & 0.046300 \\
M1 & Opening-TV & 22/28 & -0.017461 & -0.018883 & -0.058385 & 0.027417 \\
M2 & WDL-TV & 13/28 & 0.000241 & 0.001650 & -0.018300 & 0.015700 \\
M2 & Opening-TV & 27/28 & -0.117509 & -0.105987 & -0.364176 & 0.004373 \\
M3 & WDL-TV & 19/28 & -0.011037 & -0.010500 & -0.064400 & 0.032800 \\
M3 & Opening-TV & 27/28 & -0.109967 & -0.090846 & -0.347976 & 0.006572 \\
\bottomrule
\end{tabular}
\end{table*}

For M1, the aggregate improvements are broadly reflected at the dyad level. WDL-TV decreases for 22 of 28 dyads (78.6\%), with a median $AB-GG$ change of -0.010900. Opening-family TV also decreases for 22 of 28 dyads, with a median change of -0.018883. The effect is nevertheless heterogeneous: the WDL-TV changes range from -0.088700 to 0.046300, while opening-family changes range from -0.058385 to 0.027417.

The dyad-level results for M2 further separate its two behaviours. WDL-TV improves for only 13 of 28 dyads (46.4\%), with a median change of 0.001650, consistent with the near-zero aggregate effect. In contrast, opening-family TV improves for 27 of 28 dyads (96.4\%), with a median change of -0.105987. The opening effect is therefore not explained solely by a small number of player pairs, although its magnitude varies considerably across dyads.

M3 shows an intermediate WDL pattern, improving 19 of 28 dyads (67.9\%) with a median change of -0.010500. Its opening-family result is again highly consistent across pairs, improving 27 of 28 dyads with a median change of -0.090846.

\begin{figure*}[t]
\centering
\begin{minipage}{0.49\textwidth}
\centering
\includegraphics[width=\linewidth]{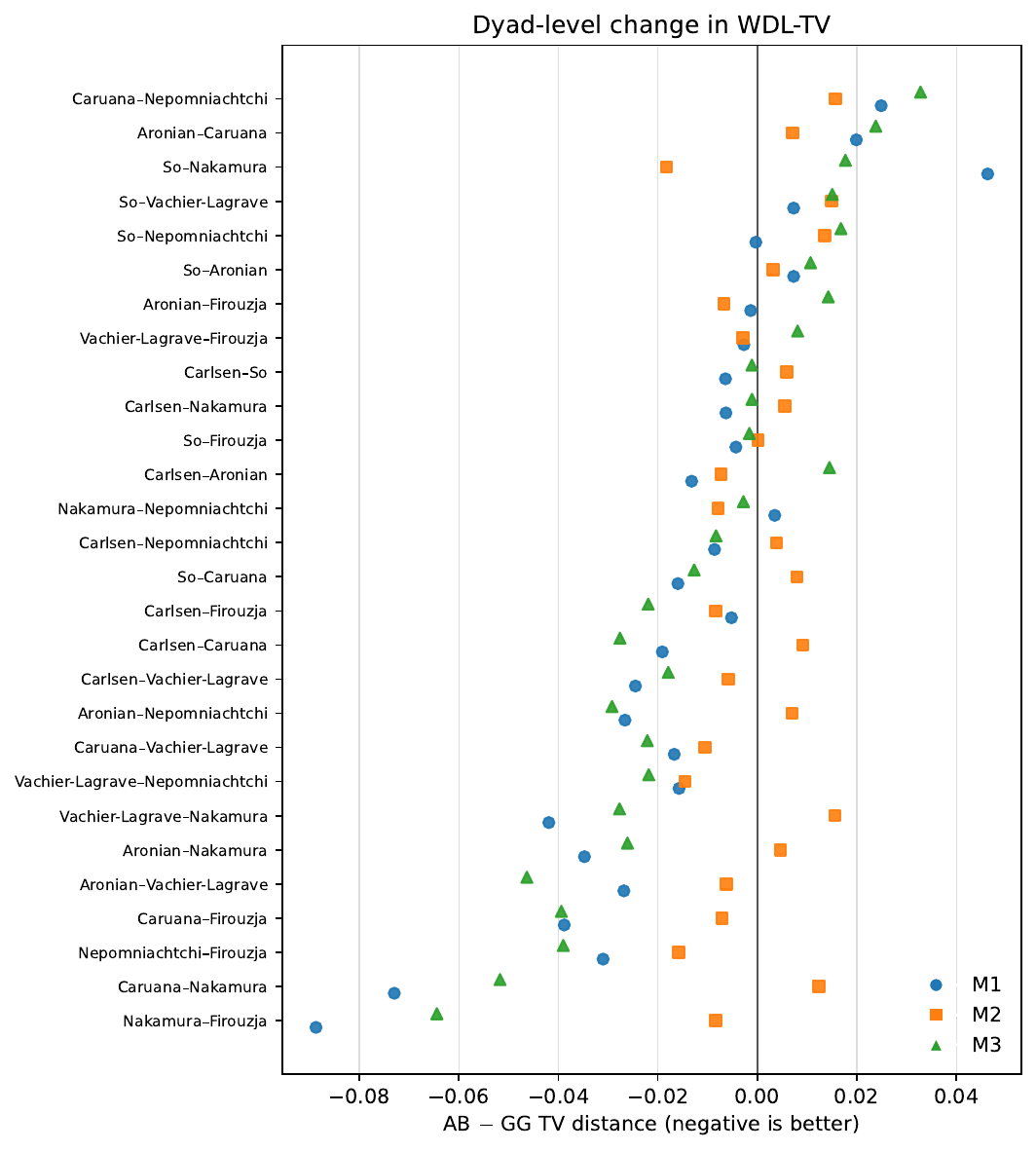}
\small (a) WDL-TV
\end{minipage}
\hfill
\begin{minipage}{0.49\textwidth}
\centering
\includegraphics[width=\linewidth]{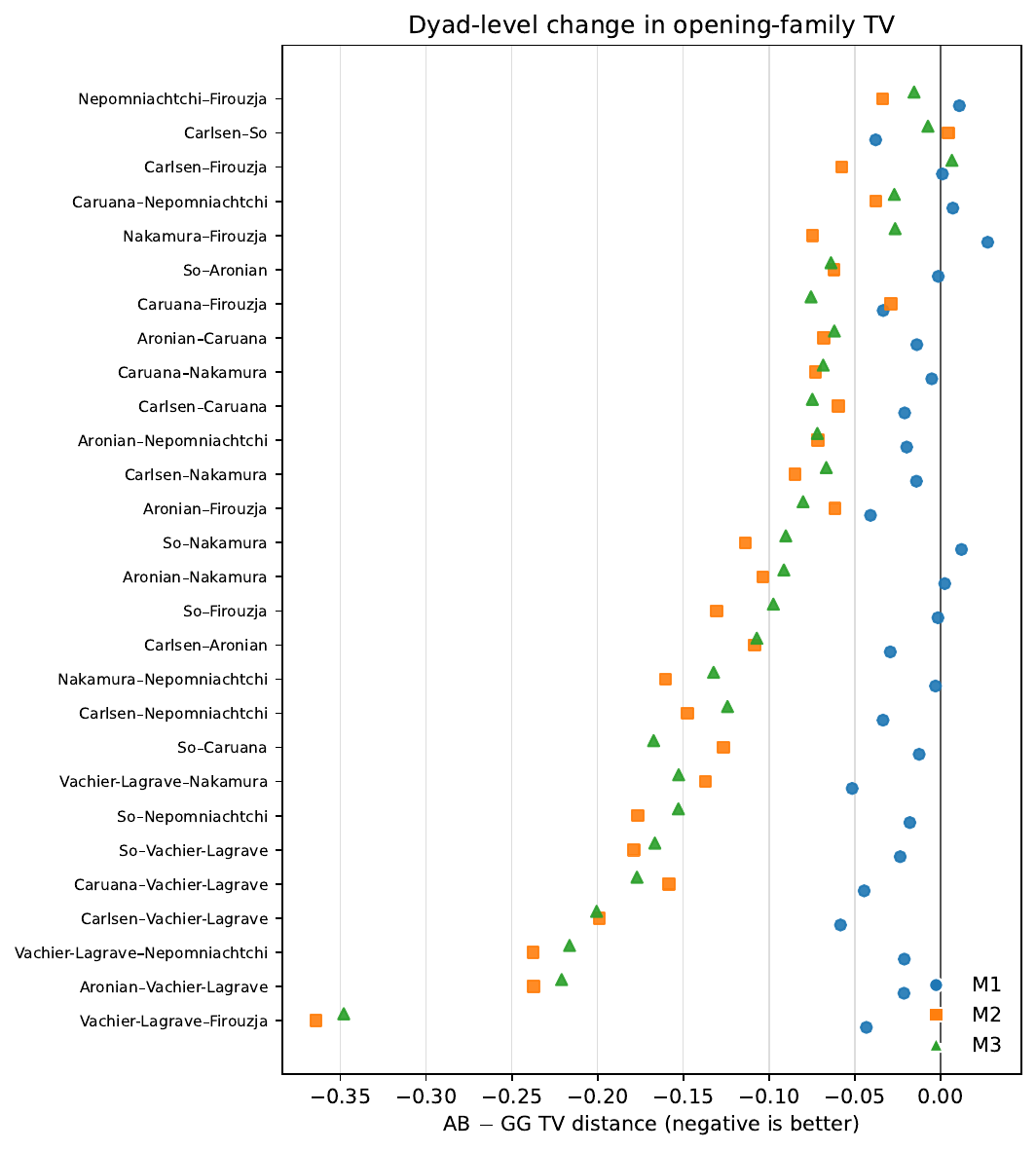}
\small (b) Opening-family TV
\end{minipage}
\caption{Dyad-level change from GG to AB for each method. Negative values indicate that the fully personalized interaction is closer to the withheld real interaction. M1 shows a broader reduction in WDL-TV, whereas the opening-family reductions for M2 and M3 are substantially larger and occur for nearly all dyads.}
\label{fig:dyad-changes}
\end{figure*}

\subsection{Identity Specificity of Compositional Recovery}

\begin{table*}[t]
\centering
\small
\caption{Identity-specificity results for the matched \AB compositions. Lower TV and rank are better; higher "Substitutions beaten" is better.}
\label{tab:identity-specificity}
\begin{tabular}{llrrrr}
\toprule
Method & Metric & Matched TV & Permutation mean & Rank / 40,320 & Substitutions beaten \\
\midrule
M1 & WDL     & 0.1501 & 0.1563 & 143   & 60\% \\
M1 & Opening & 0.5370 & 0.5460 & 1,789 & 63\% \\
M2 & WDL     & 0.1634 & 0.1643 & 8,096 & 52\% \\
M2 & Opening & 0.4372 & 0.5800 & 1     & 90\% \\
M3 & WDL     & 0.1520 & 0.1569 & 1,625 & 53\% \\
M3 & Opening & 0.4444 & 0.5886 & 1     & 90\% \\
\bottomrule
\end{tabular}
\end{table*}

We next test whether the correct learned player identities produce the closest match to the real interactions.

Table~\ref{tab:identity-specificity} shows that the strongest identity-specific effect occurs for opening-family behavior under M2 and M3. For both methods, the correct assignment of the eight learned player identities achieves the lowest mean TV of all \(8! = 40{,}320\) possible assignments. The matched M2 and M3 compositions also outperform approximately 90\% of one-player substitutions. The corresponding WDL effects are substantially weaker, particularly for M2.

\subsection{Effect of Limited Real-Game Samples}

\begin{table}[t]
  \centering
  \small
    \caption{Observed TV and finite-sample simulation results for GG and AB, averaged equally over the 28 dyads.}
  \label{tab:finite-sample}
  \begin{tabular}{lllrr}
  \toprule
  Method & Metric & Cond. & Observed TV & Reference TV \\
  \midrule
  M1 & WDL     & GG & 0.1642 & 0.1048 \\
  M1 & WDL     & AB & 0.1501 & 0.1070 \\
  M1 & Opening & GG & 0.5544 & 0.3813 \\
  M1 & Opening & AB & 0.5370 & 0.3805 \\
  \midrule
  M2 & WDL     & GG & 0.1632 & 0.1049 \\
  M2 & WDL     & AB & 0.1634 & 0.1050 \\
  M2 & Opening & GG & 0.5547 & 0.3807 \\
  M2 & Opening & AB & 0.4372 & 0.3190 \\
  \midrule
  M3 & WDL     & GG & 0.1631 & 0.1047 \\
  M3 & WDL     & AB & 0.1520 & 0.1069 \\
  M3 & Opening & GG & 0.5544 & 0.3813 \\
  M3 & Opening & AB & 0.4444 & 0.3141 \\
  \bottomrule
  \end{tabular}
\end{table}

The sealed interaction sets contain only 25--131 real games per dyad, so some TV can arise simply because the real sample is small. To estimate the size of this effect, we treat each synthetic distribution as the underlying distribution and repeatedly draw samples with the same number of games and color assignments as the corresponding real dyad.

Table~\ref{tab:finite-sample} compares the observed TV values with those produced by this finite-sample simulation. The effect is substantial for opening-family TV. For M2 and M3, the simulated TV is lower under \AB than under \GG, meaning that part of the observed \GG-to-\AB reduction coincides with the personalized opening distributions being easier to estimate from small samples. M1 shows little change in the simulated opening-family TV.

For WDL-TV, the simulated values change very little between \GG and \AB. The WDL improvements therefore do not show the same finite-sample shift seen for M2 and M3 opening-family behavior.

Across all conditions, the observed TV remains above the values produced by the finite-sample simulation. Small real-game samples therefore contribute substantially to the measured TV, particularly for opening-family behavior, but do not fully account for the gap between the synthetic and real interaction distributions.
\section{Discussion}
\label{sec:discussion}

Looking closely at the results from the previous section, we see that the different methods recover some aspects of interaction fidelity. M1 achieves better WDL fidelity than the compositions involving the generic model, while M2 results in a substantial improvement in opening-family fidelity, and M3 retains some of both. The identity-specificity analyses further show that the opening-family improvements under M2 and M3 depend strongly on composing the correct learned player identities, with the correct assignment ranking first among all \(8! = 40{,}320\) possible assignments. Together, these results suggest that independently learned models can recover aspects of previously unseen interactions, but that recovery varies strongly across different modelling methods and behavioural dimensions.

We can therefore observe partial recovery of interaction fidelity, but we cannot determine how much of the remaining gap comes from imperfect individual modelling versus properties that are genuinely specific to the interaction itself. The finite-sample analysis also shows that the limited number of sealed real games contributes substantially to the measured TV, particularly for opening-family behaviour, although it does not fully account for the remaining gap. Using both stronger individual modelling methods and larger sets of repeated real interactions would make these distinctions clearer. If stronger individual models lead to substantially better interaction fidelity, this would suggest that much of the current gap comes from limitations in the individual models. If a substantial gap remains even with stronger models and denser interaction data, however, this would provide stronger evidence that some parts of an interaction are specific to the pair itself.

The scope of the current study is also limited to eight elite chess players, chess as a single interaction domain, WDL and opening-family distributions as the main behavioural measurements, and an exploratory post-hoc hybrid in M3. Natural future directions therefore include stronger individual modelling methods, larger sets of repeated pairwise interactions, a richer set of behavioural metrics that capture more aspects of an interaction, and evaluation in other domains with repeated pairwise interactions. These extensions would help clarify both how much of an interaction can be recovered from independently learned individuals and where the limits of that recovery lie.
\section{Conclusion}
\label{sec:conclusion}

This paper studies whether independently trained models of individuals can recover aspects of previously unseen pairwise interactions. We observe partial recovery of interaction fidelity across different personalization methods: M1 improves WDL fidelity over compositions involving the generic model, M2 substantially improves opening-family fidelity, and M3 retains some of both. For M2 and M3 opening-family behaviour, the correct assignment of learned player identities also gives the closest match among all \(8! = 40{,}320\) possible assignments.

These results suggest that interaction recovery is possible in part, but depends strongly on both the individual modelling method and the behavioural property being measured. This raises a broader question: how much of an interaction can ultimately be recovered from strong independently modelled individuals, and how much reflects genuinely interaction-specific structure that cannot be recovered from the participants alone?
\section{Limitations}
\label{sec:limitations}

The study considers only eight elite chess players in a single interaction domain, and evaluates interaction fidelity using only WDL and opening-family distributions. The same patterns may not hold for weaker players, more volatile populations, other behavioural measurements, or domains outside chess.

The individual models also do not perfectly reproduce the target players, making it difficult to separate errors caused by imperfect individual modelling from behaviour that is genuinely specific to a particular interaction. The models additionally include board-defined termination but not behavioural actions such as resignation or agreed draws, which may contain player- or dyad-specific signal.

The study is limited to pairwise interactions and does not consider larger groups of models. Finally, the sealed real interaction sets contain only 25--131 games per dyad. Our finite-sample analysis shows that samples of this size can produce substantial TV, particularly for opening-family distributions, limiting how precisely the underlying real interaction distributions can be estimated.

\bibliographystyle{ACM-Reference-Format}
\bibliography{references}

\onecolumn

\clearpage
\appendix
\section{Data construction and target resolution}
\label{app:data}

\paragraph{Target cohort.}
The study concerns eight professional grandmasters (Table~\ref{tab:elo}). Each player is resolved in Lichess Broadcast metadata by exact FIDE-ID matching, falling back to exact matching against a curated list of name aliases; no fuzzy matching is used. The eight-player cohort was selected, well before model development began, by an exhaustive pairwise-coverage analysis: cohorts of size 6--9 drawn from candidate players of comparable prominence were compared, scoring each cohort by its minimum number of mutual broadcast games over any pair within it (counting only deduplicated standard games, excluding Chess960/Freestyle), and size~8 gave the best trade-off between coverage and cohort size. The production data pipeline takes this fixed eight-player list as given.

\paragraph{Game validity filter.}
The same validity filter is applied wherever games are ingested: the population corpus, the Broadcast domain-adaptation corpus, and sealed-dyad extraction. A game is kept only if
(i)~its variant is Standard (assumed when the variant header is missing);
(ii)~it does not specify a custom starting position;
(iii)~both players' Elo ratings are present and parse as integers (no plausibility bound is applied, so an Elo of~0 is accepted);
(iv)~its result is a White win, a Black win, or a draw; and
(v)~it contains at least one move.

\paragraph{Population pretraining corpus.}
The population corpus consists of five Lichess Standard-Rated monthly archives: June 2017, 2018, 2019, 2020, and 2021. Holding the calendar month fixed across years provides a consistent temporal sampling window and avoids introducing month-of-year variation as an additional source of corpus variation. Games are balanced by Elo into 22 bins keyed on the mean of the two players' ratings. One bin covers mean Elo below 600, twenty 100-point bins span $[600, 2600)$, and one bin covers mean Elo of 2600 and above. At most 10 games are kept per bin. \emph{This cap is enforced separately within each sequential chunk of 20{,}000 raw games from the source archive, not over the whole month}: every bin count resets at the start of each new 20{,}000-game chunk.

A game is also excluded if either player's Lichess username matches, after case and whitespace normalization, a curated list of the target players' known Lichess accounts. This check runs before any other processing, so no training record is ever built from an excluded game.

\paragraph{Time-pressure filtering and position sampling.}
Each game has a single cutoff for time pressure. Plies are scanned in order. The pre-move clock of a ply is the same player's clock reading two plies earlier. The first ply whose pre-move clock is strictly below 30.0\,s is ineligible for position sampling, and so is every later ply in that game. The first two plies of a game are always eligible, since no earlier clock reading exists for either color. A missing clock annotation is never taken as evidence of time pressure and never triggers the cutoff. On each training pass over the corpus, up to 32 eligible plies per game are sampled uniformly at random without replacement; a game with fewer eligible plies contributes all of them.

\paragraph{Broadcast domain-adaptation corpus.}
The domain-adaptation corpus is the Lichess Broadcast corpus of professional and elite games, with a stricter exclusion rule than the population corpus: a game is removed if \emph{either} side resolves to one of the eight targets, not only when both sides are targets.

\paragraph{Sealed dyad extraction.}
A game is \emph{sealed} if and only if its two players resolve to two \emph{distinct} members of the cohort. Sealed games are extracted from the raw Broadcast PGN corpus with the same identity-resolution and validity rules described above, then deduplicated by an exact content signature (the same two players, date, result, and full move sequence). This collapses copies of one real game published under several broadcast entries, keeping one canonical copy per duplicate group. The sealed cache used by the final evaluator contains 1{,}609 deduplicated games, covering all 28 dyads, with 25 to 131 games per dyad.

\paragraph{Leakage prevention.}
Games are classified by player identity before any training record is built. Target-vs-target (sealed) games are removed at that classification step, so they are excluded from all personalization training and validation data. No sealed game can therefore enter either player's individual training or validation data, for its own dyad or for any other.

\section{Base-model implementation and training}
\label{app:base}

\paragraph{Input representation.}
The base model is a transformer policy/value network (``Chessformer'') that reads a canonicalized board history. Positions are canonicalized so that the side to move is always presented as White; when Black is to move, the board is mirrored. The input is the current position and its 7 preceding positions. Each position is encoded as a 12-channel one-hot piece-occupancy plane per square, so each square has 96 board features. The mover's Elo and the opponent's Elo each receive a 128-dimensional embedding, computed as a linear interpolation between two learned endpoint embeddings, with Elo clamped to $[0, 5000]$. Both embeddings are concatenated with the board features of every square before the input projection, giving $96 + 128 + 128 = 352$ input features per square.

\paragraph{Backbone and heads.}
The backbone has 8 transformer blocks with model dimension 1024, 32 attention heads, and feed-forward dimension 2048. A geometric bias term, computed from a compact learned representation of board geometry shared across all 8 blocks, is added to the pre-softmax attention logits of every block. The policy head outputs 4{,}352 move logits: 4{,}096 for ordinary from-square/to-square moves and 256 for promotions (four promotion pieces for each of the $8 \times 8$ possible promoting moves). The value head outputs a three-way win/draw/loss prediction from the perspective of the side to move. Dropout is disabled (0.0) throughout; each transformer block uses RMSNorm with post-norm residual connections, while the final encoder normalization, the geometric bias module, and the value head use LayerNorm. The model has about 77.8M trainable parameters; the label ``79M'' is used informally elsewhere as a rounded figure.

\paragraph{Training.}
Training has two stages (Table~\ref{tab:base-train}). Both stages use AdamW with a learning-rate schedule of linear warmup followed by cosine annealing with restarts, automatic mixed precision, gradient-norm clipping at 3.5, and a combined policy-plus-value loss with the value term weighted at 0.1.

Stage~1 (population pretraining) trains from a random initialization on the population corpus of Section~\ref{app:data}, with target-cohort exclusion applied. During training, each sampled position's board history is independently truncated with probability 0.05 (a uniformly chosen number of preceding boards is dropped, always keeping the current board), re-rolled independently every time that position is sampled. Stage~1 uses no held-out validation split, so no checkpoint is selected by validation; the checkpoint carried forward is the one saved at the final optimizer step.

Stage~2 (Broadcast domain adaptation) starts from that Stage-1 checkpoint and trains on the target-free Broadcast corpus without Elo-bin balancing (the same history-truncation augmentation applies). The data are split deterministically into 98\% train and 2\% validation by hashing a per-game identifier. The final checkpoint was chosen by a sweep over the saved checkpoints that minimized the same combined policy-plus-value loss on this validation split.

This one Stage-2 checkpoint is frozen and reused unchanged in every later experiment (M1, M2, and M3). No personalization method retrains or fine-tunes it.

\begin{table}[ht]
\centering
\footnotesize
\caption{Base-model training configuration. Both stages use AdamW, linear warmup with cosine-annealed restarts, mixed precision, and gradient-norm clipping at 3.5.}
\label{tab:base-train}

\setlength{\tabcolsep}{4pt}
\renewcommand{\arraystretch}{1.12}

\begin{tabularx}{\textwidth}{
    @{}
    >{\raggedright\arraybackslash}p{0.25\textwidth}
    Y
    Y
    @{}
}
\toprule
& \textbf{Stage 1 (population)}
& \textbf{Stage 2 (Broadcast)} \\
\midrule

Initialization
& Random; PyTorch defaults, with Xavier initialization for Elo/GAB embeddings only
& Stage-1 final checkpoint \\

Optimizer steps
& 1{,}000{,}000
& 200{,}000 \\

Effective batch size
& 512
& 512 \\

Peak learning rate
& $5\times10^{-5}$
& $1\times10^{-5}$ \\

Learning-rate floor
& $1\times10^{-5}$
& $2\times10^{-6}$ \\

Weight decay
& $1\times10^{-6}$
& $1\times10^{-6}$ \\

Warmup / cosine cycle
& 1{,}000 / 50{,}000 steps
& 1{,}000 / 50{,}000 steps \\

Elo-bin balancing
& Yes (per 20k-game chunk)
& No \\

Validation split
& None
& 98/2 by game hash \\

Checkpoint selection
& Final step
& Validation sweep (policy $+\,0.1\times$value loss) \\

\bottomrule
\end{tabularx}
\end{table}

\section{Personalization implementation}
\label{app:personalization}

M1 and M2 train only on each player's one-target games: games between that player and a non-target opponent. Each method first splits these games 80/20 by game into training and validation (M1 and M2 do this differently; see below), and then balances the resulting move-decisions by game phase (opening, middlegame, endgame): within each split, every player's decisions are downsampled so all three phases are represented equally, to that player's smallest available phase count. This balancing is applied independently to the training and validation portions. In both methods the base model stays frozen. Table~\ref{tab:pers-train} summarizes the configurations; the learning rate and weight decay are implementation choices, not values taken from any external source.

Each target player is additionally assigned one fixed \emph{representative Elo} (Table~\ref{tab:elo}): a single conditioning value, held constant across every dyad and condition that player appears in, used wherever the base model's ordinary Elo-conditioning mechanism is not overridden by a learned representation. It is computed as the median of that player's own recorded rating across their accepted non-cohort Broadcast games; the median is used because the Broadcast corpus mixes rating provenances of different, non-interchangeable scales. This representative Elo is used for the generic condition for that player, for opponent-side conditioning whenever the other side is personalized, and to initialize M1's per-player vector below.

\paragraph{M1: Policy-level personalization.}
Each target player gets one trainable 128-dimensional vector, initialized to that player's own representative-Elo embedding. When the player is to move, this vector replaces the mover's Elo embedding; the opponent's conditioning is left unchanged, so personalization is single-sided. This vector is the only trained parameter. The objective is standard cross-entropy with legal-move masking over the full 4{,}352-action policy space, with no candidate restriction and no value loss. Each player's games are split 80/20 by game (not by position) into training and validation sets, independently for each player. The eight players are trained in eight fully independent runs, yielding eight saved vectors. At inference, the learned vector conditions the same sampling procedure used for the base model (Section~\ref{app:generation}).

\paragraph{M2: candidate-level (style-residual) personalization.}
At each decision point, the candidate set is the frozen base model's top-5 legal moves by policy logit, computed with legal-move masking. During training and validation only, the human move is appended as a sixth candidate whenever it is not already in the top~5, so the objective is always defined. At inference, the candidate set is strictly the raw top~5, so a human move outside it cannot be selected by M2.

Each candidate move is encoded by a small convolutional network whose input is 12 board-occupancy planes plus move-specific planes for the from-square, the to-square, and any promotion. Each player $p$ has a learned 32-dimensional style vector $s_p$. For candidate $m$ with feature $\phi(m)$ and frozen base logit $z_m$, the score is

\begin{equation}
  \tilde z_m
  =
  z_m
  +
  \sigma\sqrt{d}\,
  \cos\!\big(\phi(m), s_p\big),
  \qquad d = 32,
\end{equation}

where $\cos$ is cosine similarity between the $\ell_2$-normalized vectors, $d$ is the style dimension, and $\sigma$ is a single trainable scalar (initialized to 0.17) shared across all candidates and players. The trained parameters are the candidate-feature network, the style-vector table, and $\sigma$. The base model, including its value head, is frozen and serves only to produce the candidate logits.

Unlike M1, all eight players are trained \emph{jointly} in a single run, with one shared candidate network and one style-vector table. The pooled one-target data are split 80/20 into training and validation once, jointly: a single pseudo-random split state is consumed across all eight players in a fixed (sorted) order, rather than each player drawing an independent split as M1 does. The resulting split is later partitioned by player for reporting. Because the split state is shared and consumed in order, only the first player processed receives exactly the same train/validation game membership under M1 and M2; later players' splits diverge between the two methods (Section~\ref{app:individual}). The reported checkpoint is the one with the lowest joint validation loss, reached at epoch~4 in the archived run. Raw top-5 coverage of the target move was also logged as a diagnostic but was not used for selection. At inference, the raw top-5 candidates are re-scored by the style residual, and a move is sampled from the result (Section~\ref{app:generation}).

\begin{table}[ht]
\centering
\footnotesize
\caption{Personalization training configuration. The base model is frozen in both methods.}
\label{tab:pers-train}

\setlength{\tabcolsep}{4pt}
\renewcommand{\arraystretch}{1.12}

\begin{tabularx}{\textwidth}{
    @{}
    >{\raggedright\arraybackslash}p{0.25\textwidth}
    Y
    Y
    @{}
}
\toprule
& \textbf{M1}
& \textbf{M2} \\
\midrule

Trained parameters
& One 128-d vector per player
& Candidate CNN, 32-d style vector per player, scale $\sigma$ \\

Action space
& Full 4{,}352, legal-masked
& Base top-5; human move appended during train/validation if absent \\

Loss
& Cross-entropy
& Cross-entropy over candidates \\

Training scope
& 8 independent runs
& 1 joint run over all players \\

Train/val split
& 80/20 by game, independently per player
& 80/20 by game with shared seeded split state \\

Phase balancing
& Per player, per split
& Per player, per split \\

Optimizer
& AdamW
& AdamW \\

Learning rate / weight decay
& $10^{-3}$ / $10^{-4}$
& $10^{-3}$ / $10^{-4}$ \\

Batch size
& 32
& 32 \\

Max epochs / patience
& 100 / 5
& 100 / 5 \\

Checkpoint selection
& Lowest per-player validation loss
& Lowest joint validation loss \\

\bottomrule
\end{tabularx}
\end{table}

\paragraph{M3: exploratory post-hoc composition.}
M3 combines the already-trained M1 and M2 artifacts at generation time only. It adds no training, no learned parameters, and no hyperparameters, so we treat it as exploratory and not as a third method on the same footing as M1 and M2. On a personalized player's turn, the candidate logits come from that player's M1-personalized policy (the frozen base model with the player's 128-dimensional vector substituted in, exactly as in M1) instead of from the plain base model. M2's top-5 selection and style-residual reranking are then applied unchanged on top of those logits, with the same player's M2 style vector. For the generic condition, M3 uses the plain frozen base model with no M2 machinery, matching how the generic condition is defined for M1 and M2.


\section{Individual-behavior validation protocol}
\label{app:individual}

This diagnostic asks whether each learned player representation (M1's vector or M2's style vector) predicts that player's own held-out moves better than (a)~a generic, non-personalized baseline and (b)~the representations learned for the other players.

\paragraph{Data and caveat.}
The evaluation positions are the phase-balanced validation splits of Section~\ref{app:personalization}. For M1 this is each player's own validation split, 47{,}136 positions summed across the eight players; for M2 it is that player's partition of the joint split, 46{,}950 positions from the same 1{,}384 validation games. The 186-position difference reflects the two splitting procedures described in Section~\ref{app:personalization} (independent per-player splits for M1 versus one shared, sorted-order split for M2), not leakage. At the level of whole games, the validation positions are disjoint from every game used to fit the personalized parameters by gradient descent. However, \emph{the same split was used for early stopping and best-checkpoint selection}. The diagnostic is therefore a post-training check on held-out validation data, and should not be read as a result on an untouched, independent test set.

\paragraph{Metrics.}
For M1 the metric is negative log-likelihood (NLL) over the full 4{,}352-action space with legal-move masking, matching M1's training objective. For M2 it is a candidate-matched NLL over the same oracle-expanded candidate set used in M2 training (the raw top~5, plus the human move if missing). \emph{The two NLLs are on different scales, one over the full action space and the other over a small candidate set, and must not be compared or combined.}

\paragraph{Protocol and result.}
For each true player's held-out positions, NLL is computed under three kinds of representation: (i)~the generic baseline; (ii)~the player's own learned representation (``correct''); and (iii)~each of the other seven players' representations (``wrong,'' averaged). We report the rank of the correct representation among all eight by mean NLL (lower is better), the gap between the correct representation and the generic baseline, and the mean gap between the correct and wrong representations. For both M1 and M2, the correct representation ranks first (lowest NLL) for all 8 of 8 players on this diagnostic.

\paragraph{Top-5 coverage (M2 only).}
As a separate diagnostic that does not depend on any learned representation, we report for each player the fraction of held-out human moves already present in the frozen base model's raw top-5 candidates---an upper bound on how often M2's inference-time candidate restriction could recover the human move. This ranges from roughly 94\% to 96\% across the eight players, and is a different quantity from the candidate-matched NLL above.

\section{Synthetic interaction generation}
\label{app:generation}

\paragraph{Design.}
For each of the 28 dyads, the four GG/AG/GB/AB conditions are generated, each separately under both color assignments, with 5{,}000 games per assignment. This gives 10{,}000 games per dyad-condition and
\[
28 \times 4 \times 2 \times 5{,}000
=
1{,}120{,}000
\]
games per method. M1, M2, and M3 use the same factorial design and game counts.

\paragraph{Move sampling.}
At each ply, the calling policy's legal-move-masked probabilities over its top-5 candidates are reweighted by a frozen, post-hoc strength guardrail, applied identically to every method (including M3) and every condition. Let $p_1,\dots,p_5$ be the candidate probabilities from the calling policy (generic, M1-personalized, or M2's style-reranked distribution) and $\ell_i \ge 0$ the centipawn loss of candidate $i$ relative to the best of those five, measured by a depth-8 Stockfish~19 search. A move is sampled from

\begin{equation}
  q_i
  =
  \operatorname{softmax}_i
  \left(
  \log p_i
  -
  \lambda\frac{\ell_i}{100}
  \right),
  \qquad
  \lambda = 2.0.
\end{equation}

Resulting probabilities are floored at $10^{-12}$ and renormalized, so no retained candidate is ever assigned exactly zero probability. This value of $\lambda$ was selected on nonsealed, held-out data, as the smallest tested value meeting a fixed human-comparable-strength criterion; on that same held-out data it measurably reduced, but did not eliminate, the cross-player behavioral differentiation signal relative to unguarded sampling. It is applied identically across every method and condition reported here. An unguarded variant---plain temperature-1.0 sampling over the full legal-move-masked policy---exists in the codebase and was used only for earlier diagnostics; all reported results use the guarded procedure above.

\paragraph{Termination.}
Games end under the ordinary rules of chess: checkmate, stalemate, insufficient material, or a claimable draw by threefold repetition or the fifty-move rule. As a safeguard, games are capped at 500 plies. A game that reaches the cap is marked \emph{censored}, not adjudicated as a draw, and censored games are excluded from the outcome denominators in evaluation (Section~\ref{app:eval}).

\paragraph{Determinism.}
Each game's seed is derived deterministically from a fixed root seed and the game's dyad, condition, color assignment, and index. M3 uses the same seed-derivation function under a distinct namespace (the dyad identifier is prefixed), so its games never share a seed with the corresponding M1 or M2 game. Consequently, stochastic move sampling is deterministic with respect to the recorded seed and is independent of batching or execution order.

\section{Evaluation implementation}
\label{app:eval}

\paragraph{Metrics.}
For each dyad and condition, two distributions are compared against the dyad's real sealed games: (i)~the win/draw/loss outcome distribution, and (ii)~the opening-family distribution. For openings, each real and generated game is assigned a top-level named opening family from a fixed, versioned reference opening database: the game is replayed from the start position, and the name attached to the deepest position along that replay that exactly matches a database entry is used. Games reaching no recognized opening are placed in an explicit ``Unknown'' category. Both metrics are total variation distances,

\begin{equation}
  \mathrm{TV}(P,Q)
  =
  \tfrac12
  \sum_x
  \left|P(x)-Q(x)\right|.
\end{equation}

\paragraph{Aggregation.}
Each metric is computed separately for the two color assignments and then averaged exactly 50/50, never reweighted to match the color imbalance in the real sealed games. The 28 per-dyad values are averaged with equal weight to give the overall summary, so a dyad with many real games counts the same as a dyad with few.

\paragraph{Uncertainty.}
Uncertainty comes from a paired bootstrap over the real sealed games only; the fixed generated sample is not resampled. For each dyad and metric, real games are resampled with replacement within each color assignment, keeping that assignment's original sample size, for 10{,}000 replicates, and a 95\% percentile interval (the 2.5th and 97.5th percentiles of the replicate distribution) is computed per dyad and metric. \emph{Confidence intervals were computed only per dyad and per metric; the aggregated 28-dyad summary has no confidence interval.}

\section{Reproducibility summary}
\label{app:repro}

\begin{table}[ht]
\centering
\small
\caption{Representative Elo values used for conditioning.}
\label{tab:elo}

\setlength{\tabcolsep}{8pt}
\renewcommand{\arraystretch}{1.08}

\begin{tabular}{lc}
\toprule
\textbf{Player} & \textbf{Representative Elo} \\
\midrule
Magnus Carlsen            & 2855 \\
Wesley So                 & 2769 \\
Levon Aronian             & 2756 \\
Fabiano Caruana           & 2786 \\
Maxime Vachier-Lagrave    & 2751 \\
Hikaru Nakamura           & 2829 \\
Ian Nepomniachtchi        & 2778 \\
Alireza Firouzja          & 2767 \\
\bottomrule
\end{tabular}
\end{table}

\begin{table}[ht]
\centering
\footnotesize
\caption{Reproducibility summary.}
\label{tab:repro}

\setlength{\tabcolsep}{5pt}
\renewcommand{\arraystretch}{1.12}

\begin{tabularx}{\textwidth}{
    @{}
    >{\raggedright\arraybackslash}p{0.31\textwidth}
    Y
    @{}
}
\toprule

\multicolumn{2}{@{}l}{\textit{Software}} \\
Python / PyTorch
& 3.12.3 / 2.6.0 (CUDA 12.4) \\

python-chess / pyarrow
& 1.11.2 / 25.0.1 \\

\midrule
\multicolumn{2}{@{}l}{\textit{Seeds}} \\

Training
& Seed 0 where specified by the training pipeline; M1/M2 data splitting and phase balancing also use seed 0 \\

Generation root seed
& 20260911; per-game seeds derived deterministically \\

Bootstrap
& 0 is the script default; the runtime value used for the final artifact is not separately recorded \\

\midrule
\multicolumn{2}{@{}l}{\textit{Training scale}} \\

Stage 1
& 1{,}000{,}000 steps, effective batch 512 \\

Stage 2
& 200{,}000 steps, effective batch 512 \\

M1 / M2
& $\leq 100$ epochs, patience 5, batch 32 \\

\midrule
\multicolumn{2}{@{}l}{\textit{Checkpoints}} \\

Base model
& Single Stage-2 checkpoint, validation-selected and frozen for all methods \\

M1
& 8 per-player vectors, selected by lowest per-player validation loss \\

M2
& 1 joint model, selected by lowest joint validation loss (epoch 4) \\

M3
& No checkpoint of its own; composes already-trained M1 and M2 components at generation time \\

\midrule
\multicolumn{2}{@{}l}{\textit{Generation and evaluation scale}} \\

Games per method
& 1{,}120{,}000 each: 5{,}000 per dyad $\times$ condition $\times$ color assignment \\

Sampling
& Guardrail with top-5 candidates, depth-8 Stockfish~19, $\lambda=2.0$; 500-ply cap with censoring \\

Real sealed games
& 1{,}609 across 28 dyads, with 25--131 games per dyad \\

Bootstrap
& 10{,}000 replicates per dyad per metric \\

\bottomrule
\end{tabularx}
\end{table}

\end{document}